\documentclass[review]{elsarticle}

\usepackage{lineno,hyperref}

\journal{Planetary and Space Science Journal}

\begin{document}

\begin{frontmatter}

\title{Comparison of Diffuse Infrared and Far-Ultraviolet emission in the Large Magellanic Cloud: The Data}


\author[mainaddress]{Gautam Saikia\corref{mycorrespondingauthor}}
\cortext[mycorrespondingauthor]{Corresponding author}
\ead{gautamsaikia91@gmail.com}
\author[secondaryaddress]{P. Shalima}
\author[mainaddress]{Rupjyoti Gogoi}
\author[mainaddress]{Amit Pathak}

\address[mainaddress]{Department of Physics, Tezpur University, Napaam-784028, India}
\address[secondaryaddress]{Regional Institute of Education Mysore, Mysuru, Karnataka-570001, India}

\begin{abstract}

Dust scattering is the main source of diffuse emission in the far-ultraviolet (FUV). For several locations in the Large Magellanic Cloud (LMC), \textit{Far Ultraviolet Spectroscopic Explorer (FUSE)} satellite has observed diffuse radiation in the FUV with intensities ranging from 1000 -– 3 $\times$ 10$^5$ photon units and diffuse fraction between 5$\%$ - 20$\%$ at 1100 \AA. Here, we compare the FUV diffuse emission with the mid-infrared (MIR) and far-infrared (FIR) diffuse emission observed by the \textit{Spitzer Space Telescope} and the \textit{AKARI} satellite for the same locations. The intensity ratios in the different MIR and FIR bands for each of the locations will enable us to determine the type of dust contributing to the diffuse emission as well as to derive a more accurate 3D distribution of stars and dust in the region, which in turn may be used to model the observed scattering in the FUV. In this work we present the infrared (IR) data for two different regions in LMC, namely N11 and 30 Doradus. We also present the FUV$\sim$IR correlation for different infrared bands.

\end{abstract}

\begin{keyword}
\texttt{ISM: dust, infrared: ISM, Magellanic Clouds}
\end{keyword}

\end{frontmatter}


\vspace{2cm}

\section{Introduction}

Interstellar dust grains scatter and absorb incident radiation. A combination of the two processes is called
extinction (Trumpler 1930)[1]. These dust properties are all wavelength dependent as well as dependent on the composition and sizes of the grains. The scattered component is observed as diffuse emission in the optical and ultraviolet (UV). The absorbed fraction is re-emitted as diffuse radiation in the mid-infrared (MIR) or far-infrared (FIR) depending on the dust temperature (Draine 2003)[2]. Dust also is an important agent in the fluid dynamics, chemistry, heating, cooling, and even ionization balance in some interstellar regions, with a major role in the process of star formation. Despite the importance of dust, determination of the physical properties of interstellar dust grains has been a challenging task. The infrared (IR) emission from dust depends not only on the amount of dust present, but also on the rate at which it is heated by starlight.\\

The spectral properties of the IR emission from dust allow one to infer the composition of the dust, the size distribution of the dust particles, the intensity of the starlight that is heating the dust, and the total mass of dust. The dust scattered UV radiation has been observed in several regions in the Galaxy and beyond by probes like \textit{GALEX}, \textit{FUSE} etc. It has been found to be associated with regions of thin sheets of material close to hot UV emitting stars (Sujatha et al. 2005)[3]. In the MIR we can observe emission from PAH molecules as well as from tiny solid grains which have sizes starting from a few tens of Angstroms. PAH molecules can be detected in the \textit{Spitzer} 8 $\mu$m band. These are associated with colder dust clouds. The small solid grains are known as VSGs (Very Small Grains) and can be excited to very high temperatures, resulting in detectable emission in the \textit{Spitzer} 24 $\mu$m band (Wu et al. 2005)[4]. This VSG emission is seen to be associated with locations close to hot UV emitting stars like HII regions, just like the dust scattered UV radiation. However, the behaviour of the PAH emission in cluster environments  has not yet been studied well (Murata et al. 2015)[5]. This is mainly due to sparse filter sampling at 8 -- 24 $\mu$m in the \textit{Spitzer Space Telescope} (Werner et al. 2004)[6]. In contrast, the Japanese \textit{AKARI} satellite (Murakami et al. 2007)[7] has continuous wavelength coverage at 2 -- 24 $\mu$m with nine photometric bands in the \textit{Infrared Camera} (\textit{IRC}; Onaka et al. 2007)[8].\\

The Magellanic Clouds are the nearest extragalactic systems and therefore offer an opportunity for the study of extragalactic abundances (Pagel et al. 1978)[9]. Four types of objects are available for studies of this sort: normal stars, supernova remnants, planetary nebulae and HII regions. The Large Magellanic Cloud (LMC) provides a nearby, ideal laboratory to study the influence of massive stars on dust properties because it has a nearly face-on orientation, mitigating the confusion and extinction along the Galactic plane. It is at a known distance, $\sim$50 kpc (Feast 1999)[10], so stars can be resolved and studied in conjunction with the interstellar gas and dust (Stephens et al. 2014)[11]. The LMC is located at a high latitude ($\sim 30^0$; Putman et al. 1998)[12] and hence it is not much affected by extinction from the Milky Way (MW) dust. For these reasons, the LMC has been targeted by every space-based IR observatory to study dust properties and calibrate dust emission as star formation indicators. The first report of \textit{Spitzer Space Telescope} observations of an LMC H II complex was made by Gorjian et al. (2004)[13] for LHA120-N206 (N206 for short; designation from Henize 1956)[14]; its dust emission was qualitatively compared with that of the Orion Nebula. Subsequently, the entire LMC was surveyed by \textit{Spitzer} in the Legacy program Surveying the Agents of a Galaxy's Evolution (SAGE; Meixner et al. 2006)[15]. More recently, the \textit{AKARI} space telescope has been instrumental in observing the LMC.\\

Oestreicher \& Schmidt-Kaler (1996)[16] states that not only dust cloud properties, but also the distribution of the dust itself is important for understanding the structure and dynamics of the LMC. The highest reddening occurs in the regions of 30 Doradus and the supershell LMC 2 where color excess E$_{B-V}$ reaches a maximum of 0.29. The lowest reddening is observed in the region of supershell LMC 4 with E$_{B-V}$ = 0.06. The HII region N11 also shows a high reddening with  E$_{B-V}$ up to 0.24.\\  

Therefore by studying locations which have observations in the FUV and the MIR, we can hope to identify the
particular grain population responsible for the observed emission. We should expect to find a better correlation between the UV and 24 $\mu$m intensities near hot O and B-type stars compared to the UV -- 8 $\mu$m correlation. Here, we compare the FUV diffuse emission with the MIR diffuse emission observed by the \textit{Spitzer Space Telescope} and the MIR and FIR diffuse emission observed by \textit{AKARI} telescope for the same locations.

\section{Observations and Data Analysis}

We have selected a list of 81 LMC locations observed by the \textit{FUSE} UV telescope as published in Pradhan et al. (2010)[17]. Among these 81 locations, 43 were available in the \textit{Spitzer} and \textit{AKARI} archives and have been considered for this work. These 43 locations include 15 \textit{Spitzer} observations (8 $\mu$m and 24 $\mu$m) and 28 \textit{AKARI} observations (15 $\mu$m, 24 $\mu$m and 90 $\mu$m). The 8 $\mu$m data have been taken from observations by \textit{Spitzer Infrared Array Camera (IRAC)} which is a four-channel infrared camera that provides simultaneous images at four wavelengths 3.6 $\mu$m, 4.5 $\mu$m, 5.8 $\mu$m and 8 $\mu$m. All four detector arrays in the camera are 256 $\times$ 256 pixels in size, with a pixel size of $1.2^{\prime\prime}\times 1.2^{\prime\prime}$. We have taken 15 of the 43 locations containing 24 $\mu$m data from observations by \textit{Multiband Imaging Photometer for Spitzer (MIPS)} which produced imaging and photometry in three broad spectral bands in the FIR (128 $\times$ 128 pixels at 24 $\mu$m, 32 $\times$ 32 pixels at 70 $\mu$m, 2 $\times$ 20 pixels at 160 $\mu$m).\\

The 15 $\mu$m data have been taken from \textit{AKARI Infrared Camera (IRC)} which makes observations using three independent camera systems: \textit{NIR} (1.7 -- 5.5 $\mu$m), \textit{MIR-S} (5.8 -- 14.1 $\mu$m) and \textit{MIR-L} (12.4 -- 26.5 $\mu$m). We have used \textit{AKARI MIR-L} (12.4 –- 26.5 $\mu$m) camera for 15 and 24 $\mu$m observations. This camera takes data in multiple wavelengths in the 12.4 -– 26.5 $\mu$m range from which we have selected only 15 $\mu$m and 24 $\mu$m. The same camera has been used for all locations in Table \ref{intensities_akari}. $I_{15}$ refers to intensity at 15 $\mu$m wavelength, similarly for $I_8$, $I_{24}$ and $I_{90}$. We have taken 28 locations having the 24 $\mu$m data from \textit{AKARI IRC}. The 24 $\mu$m band of \textit{Spitzer} (128 $\times$ 128 pixels) has a different resolution as compared to the \textit{AKARI} (256 $\times$ 256 pixels) 24 $\mu$m and the concerned locations are different as we did not find any location with overlapping \textit{Spitzer} and \textit{AKARI} 24 $\mu$m data. One of the advantages of the \textit{IRC} was that it was able to observe 10 square arcmin at a time because of large size detector arrays (512 $\times$ 412 pixels for \textit{NIR}, 256 $\times$ 256 pixels for \textit{MIR}). We have also used 28 locations observed by the \textit{AKARI Far-Infrared Surveyor (FIS)} which was the instrument chiefly intended to make an all-sky survey at far-infrared wavelengths. The \textit{FIS} had effectively four observation bands: \textit{N60} (50 -- 80 $\mu$m), \textit{WIDE-S} (60 -- 110 $\mu$m), \textit{WIDE-L} (110 -- 180 $\mu$m) and \textit{N160} (140 -- 180 $\mu$m). We have used data taken by \textit{AKARI FIS WIDE-S} centred at 90 $\mu$m and having an array format of 15 $\times$ 3 pixels. \\

\begin{figure}[!h]
\centering
\includegraphics[width=5cm,height=5cm]{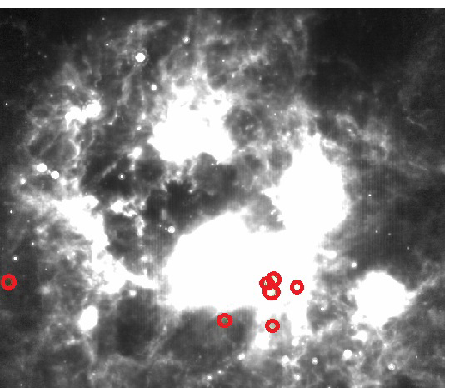}
\includegraphics[width=5cm,height=5cm]{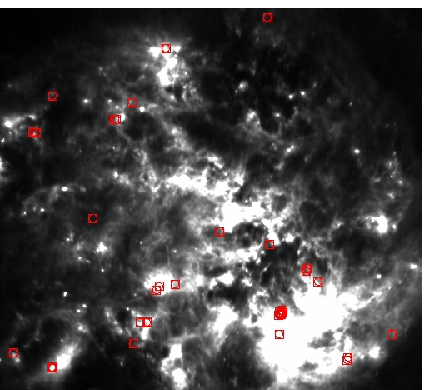}
\caption{\normalsize LMC images taken by the \textit{Spitzer Space Telescope} at 8 $\mu$m (left) showing the N11 and the \textit{AKARI} satellite at 24 $\mu$m (right) showing the 30 Doradus in the mid-IR with our locations represented in red circles and squares.}
\label{locations}
\end{figure}

Our locations were generally around well studied targets in the LMC such as the 30 Doradus (Tarantula Nebula) and N11 [Figure\ref{locations}] which are HII regions in the LMC. We have used aperture photometry technique in order to determine the intensity values for all the 43 locations and the results are tabulated in Table \ref{intensities_spitzer} and Table \ref{intensities_akari}. Table \ref{intensities_spitzer} contains 15 \textit{Spitzer} locations while Table \ref{intensities_akari} contains 28 \textit{AKARI} locations. The aperture size while calculating the intensity values has been taken as $30^{\prime\prime}\times 30^{\prime\prime}$ which is the same as the aperture size of the \textit{LWRS}, the instrument onboard the \textit{FUSE} telescope, whose data we have used in this work (Pradhan et al. 2010)[17]. The coordinates used are galactic longitude (gl) and galactic latitude (gb). The intensities ($I_8$, $I_{15}$, $I_{24}$, $I_{90}$) are in units of MJy sr$^{-1}$. The Henize numbers for the locations have been taken from the catalogue by Henize (1956)[14]. The other details have been taken from astronomical databases \textit{NED} and \textit{SIMBAD}.

\begin{center}
\begin{table}[!h]
\scriptsize
\centering
\caption{\normalsize Details of \textit{Spitzer} observations} 
\label{intensities_spitzer}
\vspace{0.5cm}
\begin{tabular}{cccccc}
\hline
\hline
Target Name & gl & gb & $I_8$ & $I_{24}$ & Henize No./Comments\\
            &    &    & (MJy sr$^{-1}$) & (MJy sr$^{-1}$) &    \\
\hline
SNR0449-693 & 280.9115 & -35.8492 & 0.90 $\pm$ 0.18 & 21.69 $\pm$ 0.48 & N77A, N77D (HII region) \\
SNR0450-709 & 282.6024 & -35.2981 & 0.21 $\pm$ 0.12 & 20.61 $\pm$ 0.09 & N78 (Emission nebula) \\
SK-67005 & 278.8562 & -36.3031 & 1.49 $\pm$ 0.11 & 20.51 $\pm$ 0.07 & N3 (Exciting star) \\
SNR0454-672 & 278.2105 & -36.0629 & 0.98 $\pm$ 0.21 & 21.07 $\pm$ 0.18 & S3 (C1 object in NGC 1735) \\
SNR0454-665 & 277.2729 & -36.2522 & 0.69 $\pm$ 0.19 & 16.89 $\pm$ 0.13 & S2 (Emission-line star) \\
SK-67D14-BKGD & 278.2475 & -36.0221 & 1.29 $\pm$ 1.03 & 16.36 $\pm$ 0.11 & S3 (C1 object in NGC 1735) \\
SNR0455-687 & 279.8756 & -35.5509 & 2.19 $\pm$ 0.16 & 16.03 $\pm$ 0.05 & N84 (Emission nebula) \\
LH103204 & 277.1267 & -36.0851 & 2.23 $\pm$ 0.21 & 17.76 $\pm$ 0.23 & N11, N11B (HII region, HD 32256) \\
PGMW-3223 & 277.1008 & -36.0451 & 4.38 $\pm$ 1.08 & 19.16 $\pm$ 0.36 & N11A (HII region, HD 32340)\\
SK-68D15  & 279.4643 & -35.5125 & 0.70 $\pm$ 0.20 & 21.04 $\pm$ 0.42 & N91 (HII region, NGC 1770) \\
LH103073 & 277.1354 & -36.0335 & 6.98 $\pm$ 2.61 & 26.97 $\pm$ 3.30 & N11A (HII region, HD 32340) \\
PGMW-3053 & 277.1473 & -36.0309 & 20.78 $\pm$ 7.73 & 60.05 $\pm$ 39.05 & N11A (HII region, HD 32340)\\
PGMW-3070 & 277.1460 & -36.0271 & 30.87 $\pm$ 24.11 & 83.59 $\pm$ 52.14 & N11A (HII region, HD 32340) \\
SK-68D16 & 279.4828 & -35.4927 & 2.32 $\pm$ 0.32 & 20.26 $\pm$ 0.39 & N91 (HII region, NGC 1770) \\
PGMW-3168 & 277.1277 & -36.0103 & 7.14 $\pm$ 1.30 & 23.45 $\pm$ 1.63 & N11A (HII region, HD 32340) \\
\hline
\end{tabular}
\end{table}
\end{center}

\begin{center}
\begin{table}[!h]
\tiny
\centering
\caption{\normalsize Details of \textit{AKARI} observations} 
\label{intensities_akari}
\vspace{0.5cm}
\begin{tabular}{ccccccc}
\hline
\hline
Target Name & gl & gb & $I_{15}$ & $I_{24}$ & $I_{90}$ & Henize No./Comments\\
            &    &    & (MJy sr$^{-1}$) & (MJy sr$^{-1}$) & (MJy sr$^{-1}$) &   \\
\hline
0509-67.5 & 278.1417 & -34.5918 & 5.84 $\pm$ 2.48 & 9.64 $\pm$ 5.36 & 6.86 $\pm$ 0.06 &	N25 (Emission nebula) \\
0519-69.0 & 279.7050 & -33.3704 & 17.72 $\pm$ 8.34 & 78.61 $\pm$ 19.20 & 19.02 $\pm$ 0.43 & S99, N120 (SNR, NGC 1918) \\
SNR0521-657 & 275.7129 & -33.7330 & 46.34 $\pm$ 12.87 & 65.15 $\pm$ 23.61 & 24.85 $\pm$ 2.40 &	N40, N43 (NGC 1923) \\
N44C-NEBULA & 278.3468 & -33.2801 & 97.19 $\pm$ 47.31 & 249.71 $\pm$ 79.21 & 153.25 $\pm$ 20.52 & N44 (HD 269404) \\
SNR0523-679 & 278.2700 & -33.2605 & 16.47 $\pm$ 7.62 & 26.39 $\pm$ 12.33 & 55.53 $\pm$ 4.91 & N44L, N44 (HII region, IC 2128) \\
Sk-68.75 & 278.5779 & -33.1640 & 5.81 $\pm$ 4.36 & 21.27 $\pm$ 10.91 & 36.55 $\pm$ 0.78 & N44L, N44 (HII region, IC 2128) \\
N132D & 280.2325 & -32.7985 & 4.93 $\pm$ 2.56 & 9.73 $\pm$ 5.31 & 15.91 $\pm$ 0.35 & N139 (Emission nebula) \\
N49-KNOT & 276.1045 & -33.2577 & 76.17 $\pm$ 26.47 & 91.11 $\pm$ 34.81 & 60.66 $\pm$ 4.69 &	N48B (NGC 1945) \\
N49-W & 276.0920 & -33.2551 & 449.14 $\pm$ 194.29 & 163.66 $\pm$ 37.50 & 69.36 $\pm$ 7.86 & N48B (NGC 1945)\\
N49 & 276.1019 & -33.2417 & 39.46 $\pm$ 13.10 & 619.49 $\pm$ 327.36 & 80.14 $\pm$ 7.53 & N49 (SNR, HD 271255) \\
SK-67D104-BKGD & 277.7323 & -33.0396 & 9.56 $\pm$ 4.45 & 21.56 $\pm$ 5.84 & 41.78 $\pm$ 1.50 & N51D (C1 object, NGC 1974) \\
SK-67.106 & 277.8186 & -32.9824 & 31.62 $\pm$ 8.17 & 60.14 $\pm$ 12.47 & 58.22 $\pm$ 2.05 & N51D (C1 object, NGC 1974)\\
SK-67.107 & 277.8173 & -32.9749 & 19.18 $\pm$ 9.46 & 32.42 $\pm$ 13.82 & 53.08 $\pm$ 4.64 & N51D (C1 object, NGC 1974) \\
SNR0528-672 & 277.4350 & -32.8499 & 8.79 $\pm$ 3.51 & 11.09 $\pm$ 5.20 & 13.47 $\pm$ 0.56 & S36 (Wolf-Rayet star), N51 \\
SK-65D63 & 275.4692 & -33.0361 & 8.19 $\pm$ 3.39 & 28.84 $\pm$ 5.69 & 4.41 $\pm$ 0.08 & \\
AADOR-BKGD & 280.4771 &  -32.2126 & 3.83 $\pm$ 1.56 & 6.44 $\pm$ 3.68 & 27.44 $\pm$ 0.62 & S117 (Blue Supergiant), N135 \\
AADOR-BKGD & 280.4365 & -32.1865 & 4.81 $\pm$ 2.46 & 8.53 $\pm$ 3.74 & 20.28 $\pm$ 0.59 & S117, N135 (NGC 2052) \\
SNR0534-699 & 280.4831 & -31.9756 & 7.63 $\pm$ 3.52 & 13.96 $\pm$ 4.23 & 34.50 $\pm$ 0.71 & N154B, N152 (C1 object, NGC 2033) \\
SNR1987A-STAR3 & 279.6733 & -31.9713 & 102.20 $\pm$ 38.83 & 250.91 $\pm$ 70.544 & 146.32 $\pm$ 14.71 & N154, N154B (C1 object, NGC 2033) \\
SN1987A& 279.7307 & -31.9578 & 12.12 $\pm$ 5.82 & 13.08 $\pm$ 6.42 & 60.25 $\pm$ 4.31 & N154, N154B(C1 object, NGC 2033) \\
SN1987A-STAR3 & 279.6334 & -31.9436 & 278.78 $\pm$ 63.10 & 621.32 $\pm$ 125.29 & 369.12 $\pm$ 59.09 & N154, N154B (C1 object, NGC 2033) \\
SN1987A & 279.6684 & -31.9397 &	86.03 $\pm$ 23.51 & 210.22 $\pm$ 32.28 & 162.81 $\pm$ 11.81 & N154, N154B (C1 object, NGC 2033) \\
SN1987A & 279.7035 & -31.9358 &	95.93 $\pm$ 15.11 & 158.55 $\pm$ 28.12 & 86.95 $\pm$ 7.99 & N154, N154B (C1 object, NGC 2033) \\
SN1987A & 279.6674 & -31.9327 &	101.77 $\pm$ 21.59 & 245.51 $\pm$ 45.39 & 187.02 $\pm$ 29.56 & N154, N154B (C1 object, NGC 2033) \\
SK-69D243 & 279.4482 & -31.7298 & 585.03 $\pm$ 232.79 & 2337.53 $\pm$ 767.45 & 586.16 $\pm$ 81.12 & N157A, N157A (HII region, HD 38268) \\
SNR0547-697B & 280.1003 & -30.8812 & 9.15 $\pm$ 4.41 & 7.79 $\pm$ 3.67 & 54.74 $\pm$ 2.01 & N170 (Emission nebula) \\
SNR0547-697A & 280.0631 & -30.8595 & 27.85 $\pm$ 7.63 & 32.92 $\pm$ 8.59 & 55.09 $\pm$ 1.98 & N170 (Emission nebula) \\
0548-70.4 & 280.8937 & -30.7529 & 7.99 $\pm$ 2.45 &	11.69 $\pm$ 5.57 & 19.43 $\pm$ 0.51 & \\  
\hline
\end{tabular}
\end{table}
\end{center}

\section{Results and Discussion}

We have calculated the rank correlations among the intensity values for FUV and the five IR wavelengths that we have observed. The rank correlation is an important statistical tool to study the relation between two quantities. Higher value of rank correlation coefficient signifies better agreement between the two quantities. The rank correlation coefficient is inside the interval [-1,1] and takes the value 1 if the agreement between the two rankings is perfect (the two rankings are the same), 0 if the rankings are completely independent and -1 if the disagreement between the two rankings is perfect (one ranking is the reverse of the other).\\

We have used the Spearman's (rho) rank correlation which is a particular case of the general correlation coefficient. For a sample of size n, the n raw scores $X_i$, $Y_i$ are converted to ranks $x_i$, $y_i$ and $\rho$ is computed from:
$$ \rho = 1 - \frac{6 \Sigma d_i^2}{n(n^2  -1)}$$
where $d_i$ = $x_i$ -– $y_i$, is the difference between ranks.\\
 
The probability or the standard error of the coefficient ($\sigma $) is given as:
$$ \sigma = \frac{0.6325}{\sqrt{(n-1)}} $$
This is the probability of seeing the observed correlation or stronger if no correlation exists. So lower the value of $\sigma$, more is the reliability in the observed value of rank correlation.\\

The far-ultraviolet (FUV) intensity values are taken from \textit{FUSE} observations (Pradhan et al. 2010)[17]. The \textit{FUSE} spacecraft included three apertures: the \textit{HIRS} ($1.25^{\prime\prime}\times 20^{\prime\prime}$), the \textit{MDRS} ($4^{\prime\prime}\times 20^{\prime\prime}$), and the \textit{LWRS} ($30^{\prime\prime}\times 30^{\prime\prime}$), all of which obtained data simultaneously. Only the \textit{LWRS}, with its relatively large field of view, was useful for diffuse observations.  Murthy $\&$ Sahnow (2004)[18] have described the analysis of the background observations, and Pradhan et al. (2010)[17] have followed their method of extraction of diffuse surface brightness from the \textit{FUSE} spectra. The \textit{FUSE} spectrum is treated as a broadband photometric observation and the spectra is collapsed into two wavelength bands per detector, excluding the terrestrial airglow lines (primarily Ly$\beta$). This results in seven wavelength bands with effective wavelengths at 1004 \AA (1A1), 1058 \AA (1A2), 1117 \AA (1B1), 1157 \AA (1B2), 1159 \AA (2A1), 1112 \AA (2A2) and 1056 \AA (2B1) (Pradhan et al. 2010)[17].\\

\begin{figure}[!h]
\centering
\includegraphics[width=8cm,height=6cm]{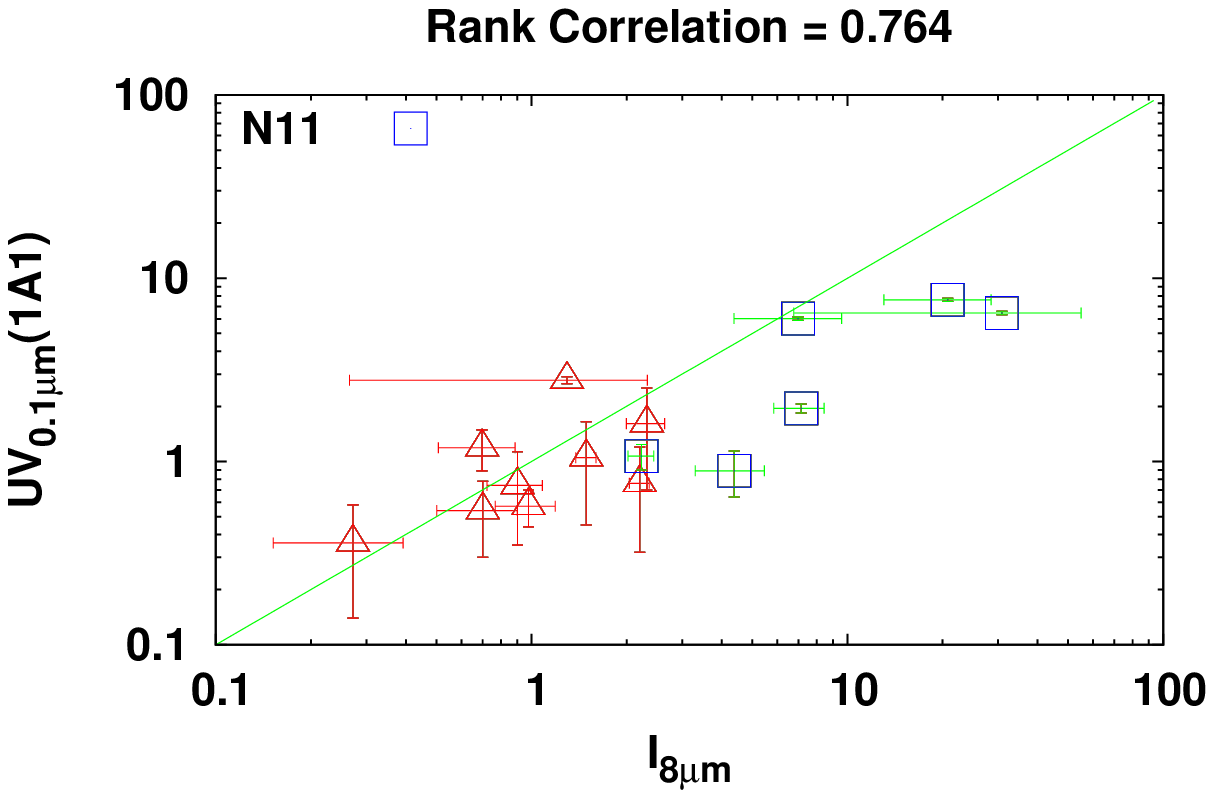}
\includegraphics[width=8cm,height=6cm]{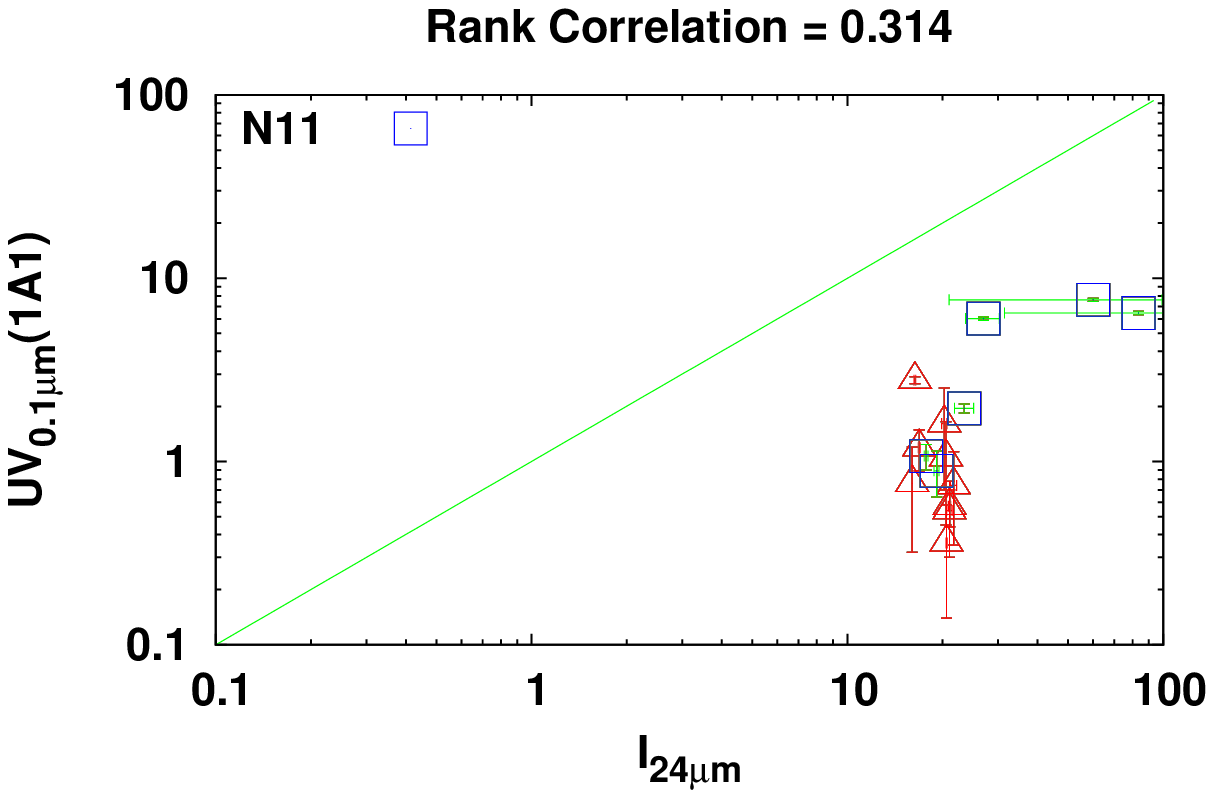}
\caption{\normalsize Correlations plotted for the 15 \textit{Spitzer} locations (Table \ref{Spitzer_all})}
\label{Spitzer_all_graphs}
\end{figure}

\begin{figure}[!h]
\centering
\includegraphics[width=8cm,height=6cm]{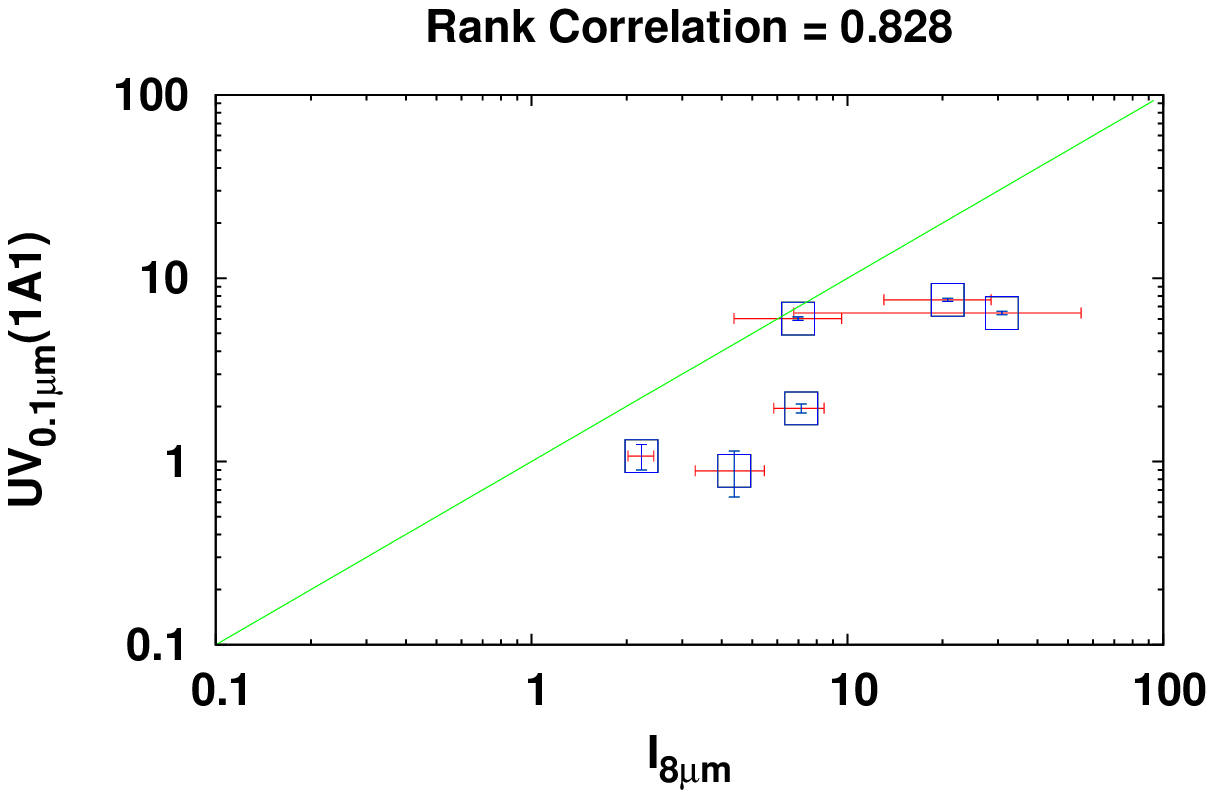}
\includegraphics[width=8cm,height=6cm]{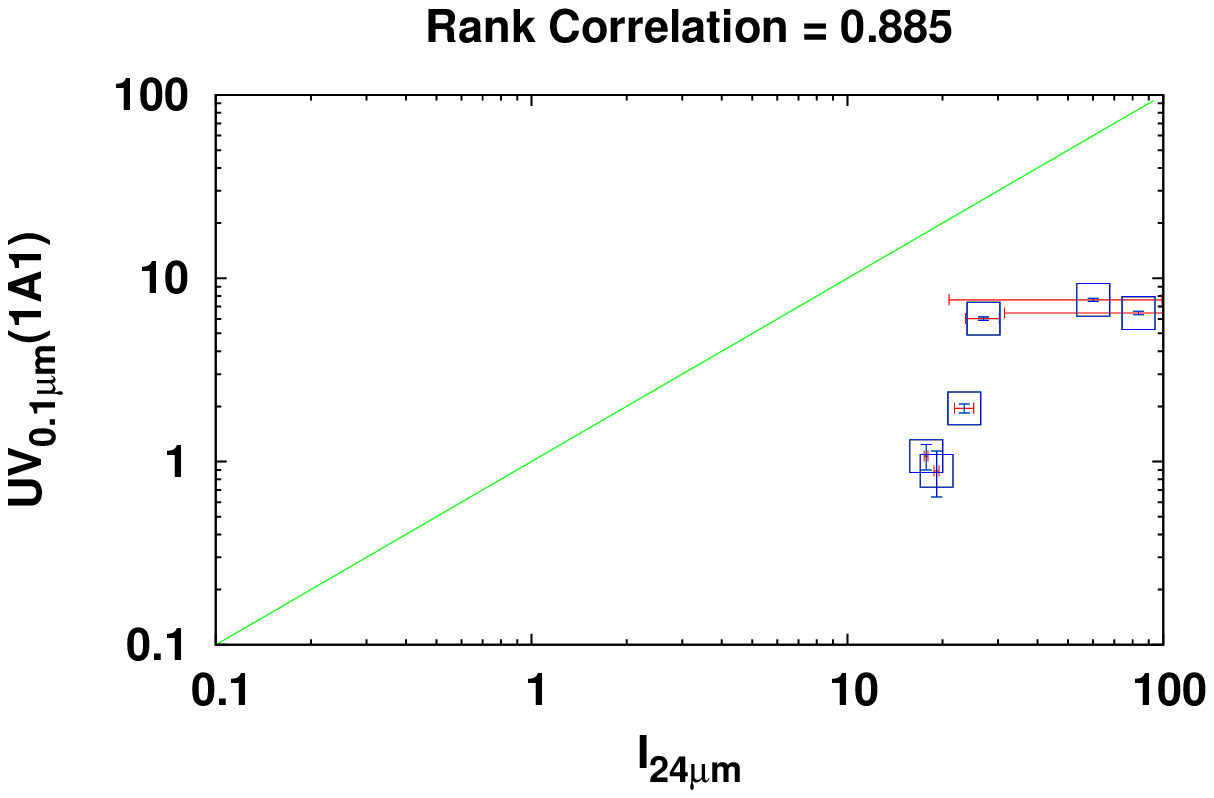}
\caption{\normalsize Correlations plotted for the N11 \textit{Spitzer} locations (Table \ref{Spitzer_N11})}
\label{Spitzer_N11_graphs}
\end{figure}

\begin{figure}[!h]
\centering
\includegraphics[width=8cm,height=6cm]{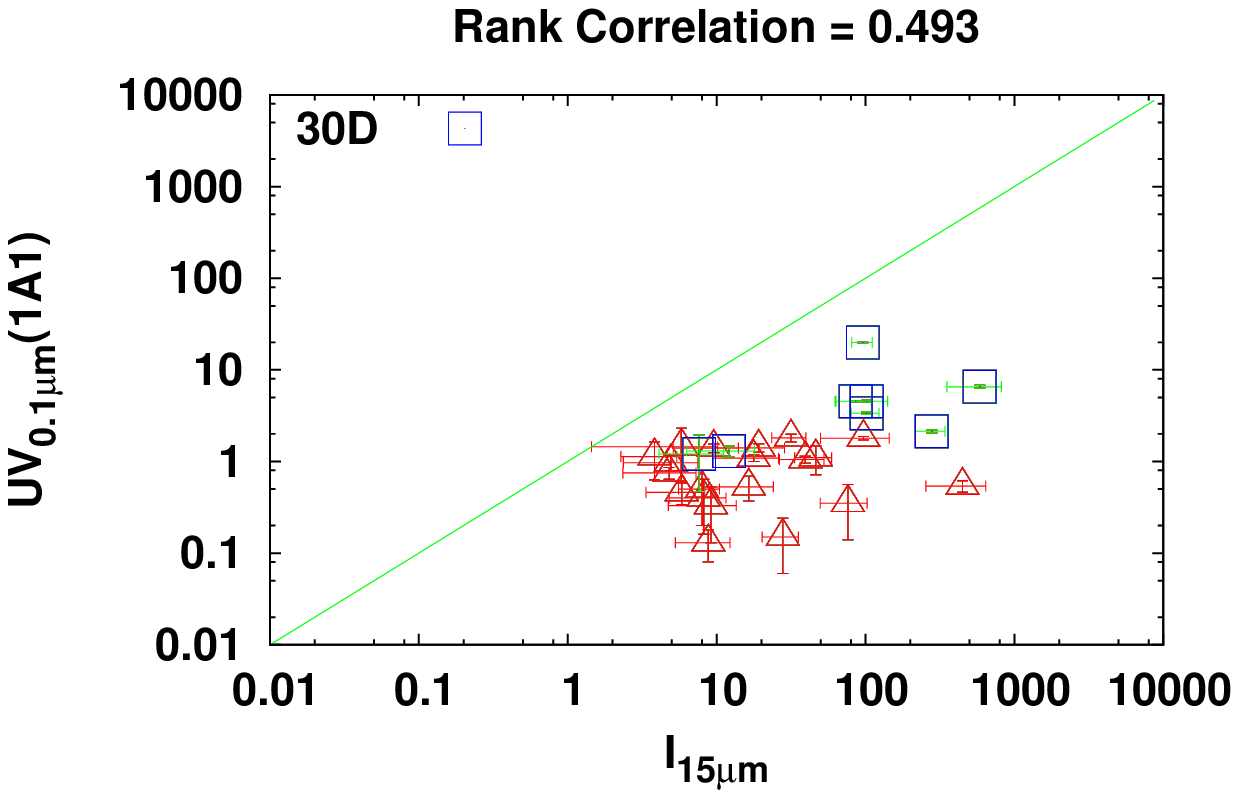}
\includegraphics[width=8cm,height=6cm]{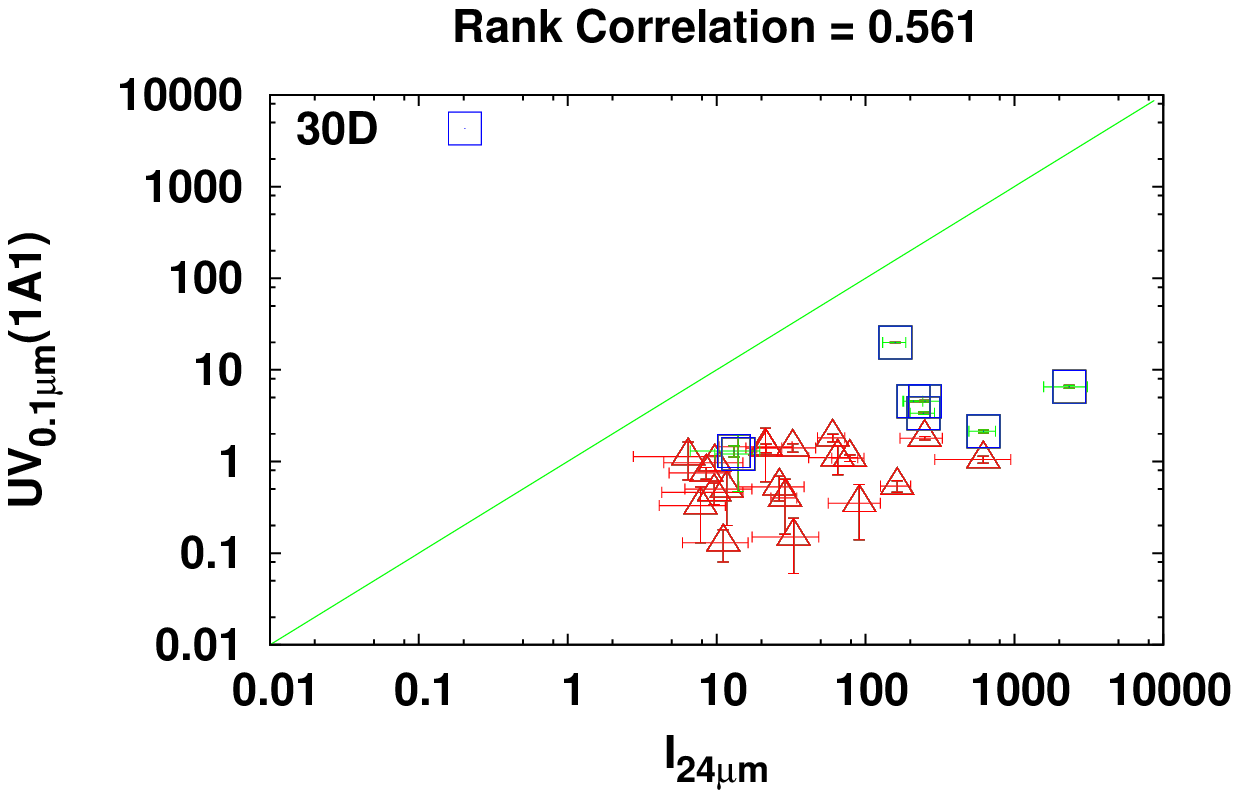}
\includegraphics[width=8cm,height=6cm]{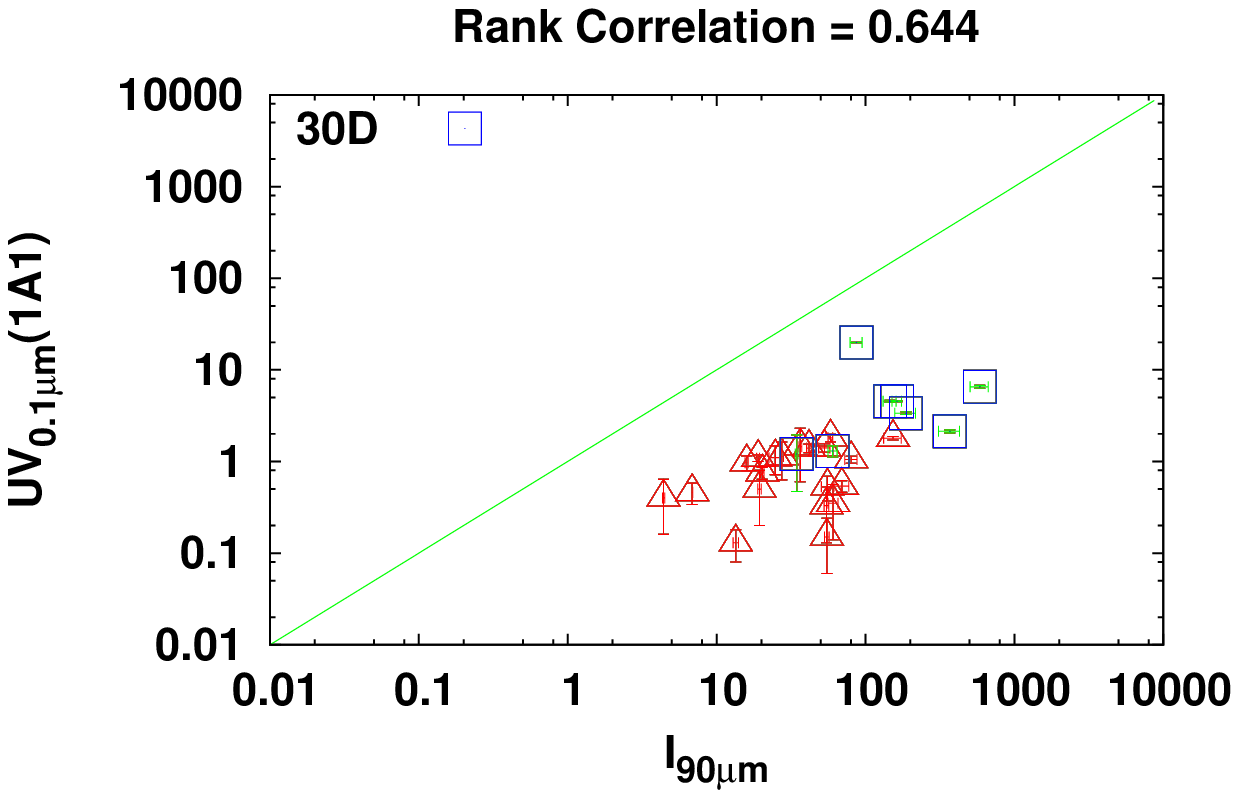}
\caption{\normalsize Correlations plotted for the 28 \textit{AKARI} locations (Table \ref{Akari_all})}
\label{Akari_all_graphs}
\end{figure}

\begin{figure}[!h]
\centering
\includegraphics[width=8cm,height=6cm]{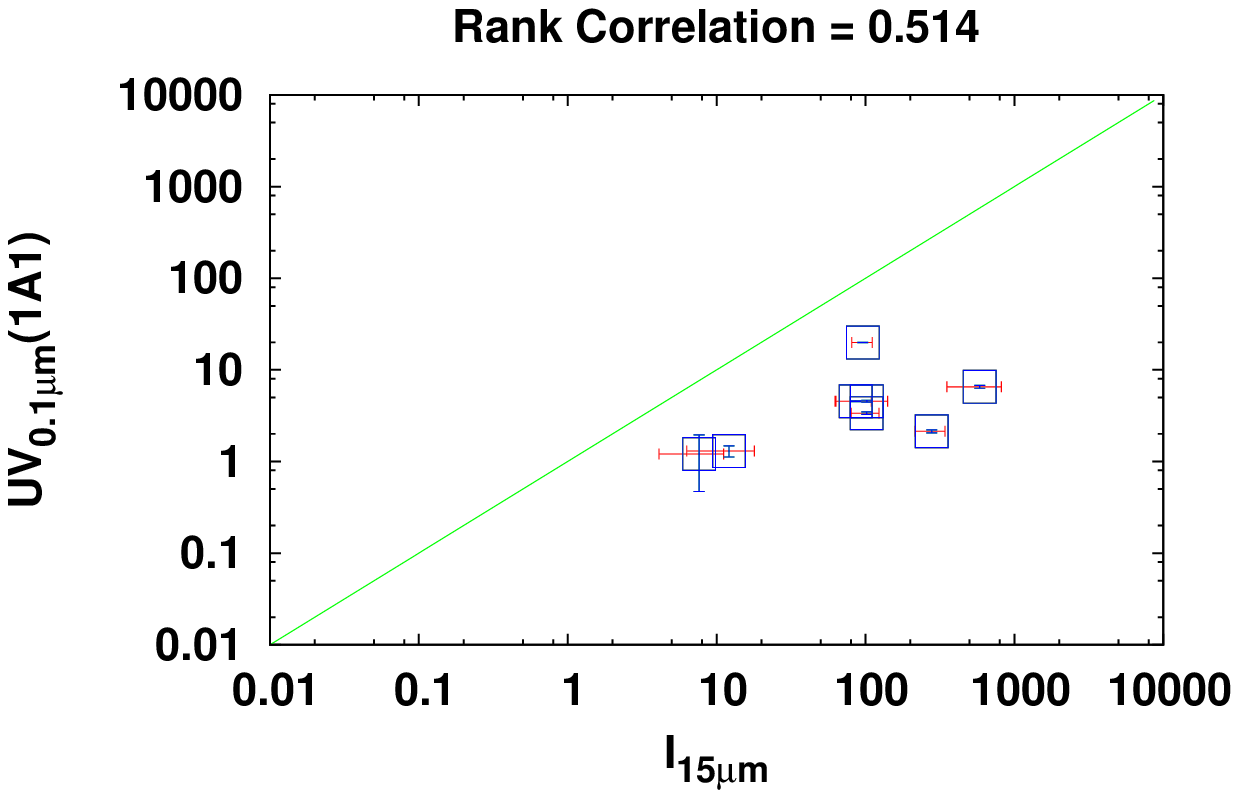}
\includegraphics[width=8cm,height=6cm]{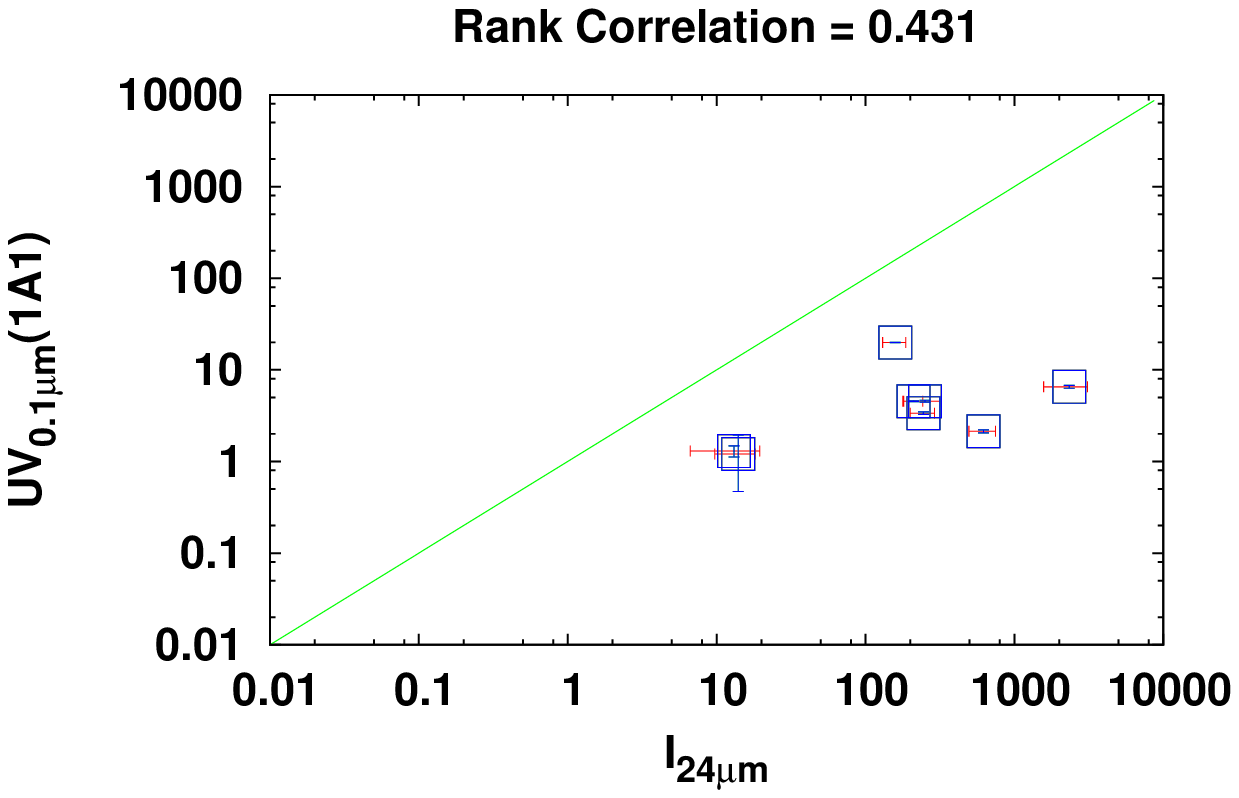}
\includegraphics[width=8cm,height=6cm]{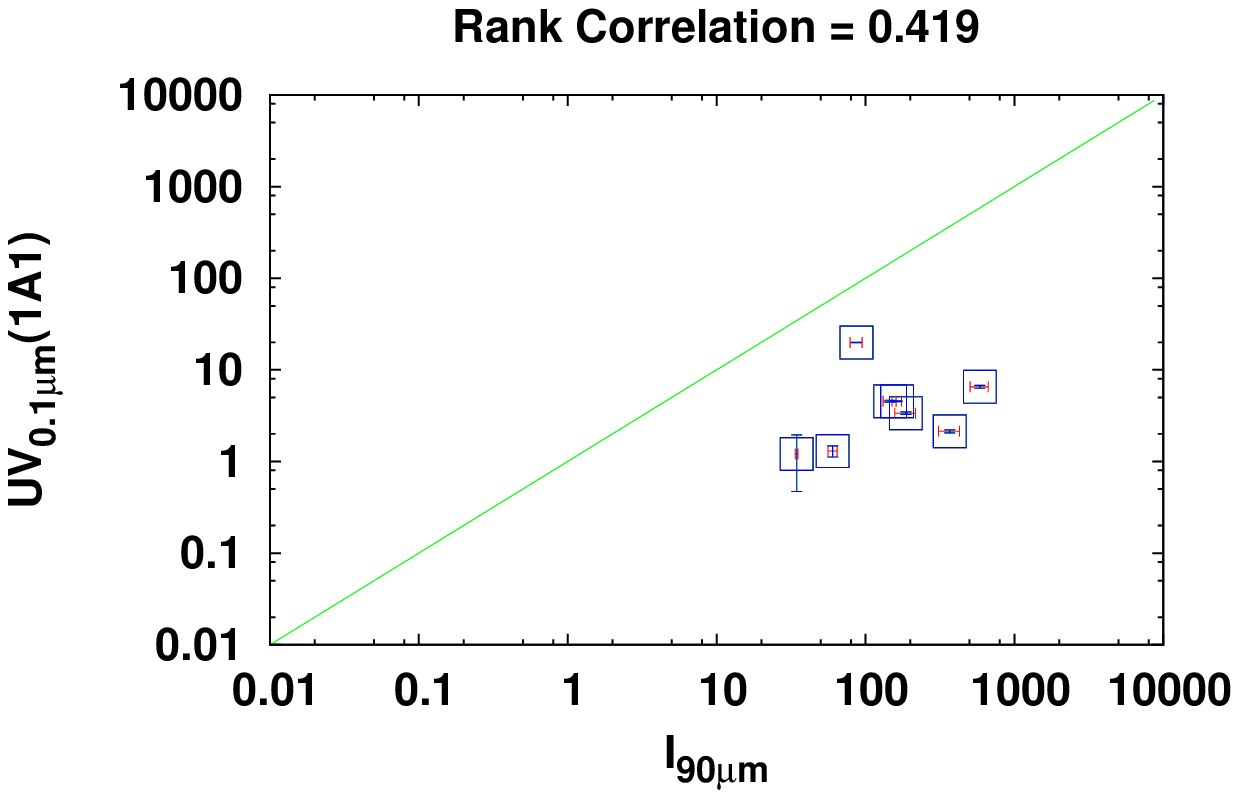}
\caption{\normalsize Correlations plotted for the 30 Doradus \textit{AKARI} locations (Table \ref{Akari_30D})}
\label{Akari_30D_graphs}
\end{figure}

The correlations for \textit{Spitzer} IR data at 8 $\mu$m and 24 $\mu$m with \textit{FUSE} FUV are shown in Table \ref{Spitzer_all}. The corresponding \textit{Spitzer-FUSE} correlation graphs are shown in Figure \ref{Spitzer_all_graphs}. We have also calculated the correlations separately for N11 and the results are presented in Table \ref{Spitzer_N11}. The corresponding N11 correlation graphs are shown in Figure \ref{Spitzer_N11_graphs}. The correlations for \textit{AKARI} IR data at 15 $\mu$m, 24 $\mu$m and 90 $\mu$m with \textit{FUSE} FUV are shown in Table \ref{Akari_all}. The corresponding \textit{AKARI-FUSE} correlation graphs are shown in Figure \ref{Akari_all_graphs}. We have also calculated the correlations separately for the 30 Doradus region in the LMC and the results are given in Table \ref{Akari_30D}. The corresponding 30 Doradus graphs are shown in Figure \ref{Akari_30D_graphs}.\\

\begin{center}
\begin{table}[!h]
\scriptsize
\centering
\caption[]{\normalsize Correlations for all 15 \textit{Spitzer} locations} 
\label{Spitzer_all}
\begin{tabular}{ccc}
\hline
\hline
IR-FUV & Rank correlation & Probability\\
\hline
$I_8$ $\sim$ fuv1A1  &   0.764 &   0.001\\
$I_8$ $\sim$ fuv1A2  &   0.610  &  0.015\\
$I_8$ $\sim$ fuv1B1  &   0.752  &  0.001\\
$I_8$ $\sim$ fuv1B2  &   0.739  &  0.002\\
$I_8$ $\sim$ fuv2A1  &   0.817  &  0.000\\
$I_8$ $\sim$ fuv2A2  &   0.696  &  0.004\\
$I_8$ $\sim$ fuv2B1  &   0.682  &  0.005\\
\hline
$I_{24}$ $\sim$ fuv1A1  &   0.314  &   0.254\\
$I_{24}$ $\sim$ fuv1A2  &   0.357  &   0.191\\
$I_{24}$ $\sim$ fuv1B1  &   0.439  &   0.101\\
$I_{24}$ $\sim$ fuv1B2  &   0.464  &   0.081\\
$I_{24}$ $\sim$ fuv2A1  &   0.442  &   0.098\\
$I_{24}$ $\sim$ fuv2A2  &   0.478  &   0.071\\
$I_{24}$ $\sim$ fuv2B1  &   0.485  &   0.066\\
\hline
\end{tabular}
\end{table}
\end{center}

\begin{center}
\begin{table}[!h]
\scriptsize
\centering
\caption[]{\normalsize Correlations for the 6 `N11' \textit{Spitzer} locations} 
\label{Spitzer_N11}
\begin{tabular}{ccc}
\hline
\hline
IR-FUV & Rank correlation & Probability\\
\hline
$I_8$ $\sim$ fuv1A1  &   0.828  &  0.041 \\
$I_8$ $\sim$ fuv1A2  &   0.657  &  0.156 \\
$I_8$ $\sim$ fuv1B1  &   0.657  &  0.156 \\
$I_8$ $\sim$ fuv1B2  &   0.714  &  0.110 \\
$I_8$ $\sim$ fuv2A1  &   0.885  &  0.018 \\
$I_8$ $\sim$ fuv2A2  &   0.828  &  0.041 \\
$I_8$ $\sim$ fuv2B1  &   0.885  &  0.018 \\
\hline
$I_{24}$ $\sim$ fuv1A1  &   0.885  &  0.018 \\
$I_{24}$ $\sim$ fuv1A2  &   0.771  &  0.072 \\
$I_{24}$ $\sim$ fuv1B1  &   0.771  &  0.072 \\
$I_{24}$ $\sim$ fuv1B2  &   0.828  &  0.041 \\
$I_{24}$ $\sim$ fuv2A1  &   0.942  &  0.004 \\
$I_{24}$ $\sim$ fuv2A2  &   0.885  &  0.018 \\
$I_{24}$ $\sim$ fuv2B1  &   0.942  &  0.004 \\
\hline
\end{tabular}
\end{table}
\end{center}

\begin{center}
\begin{table}[!h]
\scriptsize
\centering
\caption[]{\normalsize Correlations for all 28 \textit{AKARI} locations} 
\label{Akari_all}
\begin{tabular}{ccc}
\hline
\hline
IR-FUV & Rank correlation & Probability\\
\hline
$I_{15}$ $\sim$ $I_{24}$  &   0.918 & 5.64e-12 \\
$I_{15}$ $\sim$ $I_{90}$  &   0.834 & 3.35e-08 \\
$I_{24}$ $\sim$ $I_{90}$  &   0.779 & 1.02e-06 \\
\hline
$I_{15}$ $\sim$ fuv1A1  &   0.493 & 0.007 \\
$I_{15}$ $\sim$ fuv1A2  &   0.586 & 0.001 \\
$I_{15}$ $\sim$ fuv1B1  &   0.668 & 0.000 \\
$I_{15}$ $\sim$ fuv1B2  &   0.630 & 0.000 \\
$I_{15}$ $\sim$ fuv2A1  &   0.574 & 0.001 \\
$I_{15}$ $\sim$ fuv2A2  &   0.626 & 0.000 \\
$I_{15}$ $\sim$ fuv2B1  &   0.571 & 0.001 \\
\hline
$I_{24}$ $\sim$ fuv1A1  &   0.561 & 0.001 \\
$I_{24}$ $\sim$ fuv1A2  &   0.657 & 0.000 \\
$I_{24}$ $\sim$ fuv1B1  &   0.707 & 2.544e-05 \\
$I_{24}$ $\sim$ fuv1B2  &   0.687 & 5.288e-05 \\
$I_{24}$ $\sim$ fuv2A1  &   0.639 & 0.000 \\
$I_{24}$ $\sim$ fuv2A2  &   0.658 & 0.000 \\
$I_{24}$ $\sim$ fuv2B1  &   0.596 & 0.001 \\
\hline
$I_{90}$ $\sim$ fuv1A1  &   0.644 & 0.000 \\
$I_{90}$ $\sim$ fuv1A2  &   0.620 & 0.000 \\
$I_{90}$ $\sim$ fuv1B1  &   0.753 & 3.674 \\
$I_{90}$ $\sim$ fuv1B2  &   0.724 & 1.31e-05 \\
$I_{90}$ $\sim$ fuv2A1  &   0.697 & 3.70e-05 \\
$I_{90}$ $\sim$ fuv2A2  &   0.709 & 2.37e-05 \\
$I_{90}$ $\sim$ fuv2B1  &   0.672 & 8.79e-05 \\
\hline
\end{tabular}
\end{table}
\end{center}

\begin{center}
\begin{table}[!h]
\scriptsize
\centering
\caption[]{\normalsize Correlations for the 8 `30 Doradus' \textit{AKARI} locations} 
\label{Akari_30D}
\begin{tabular}{ccc}
\hline
\hline
IR-FUV & Rank correlation & Probability\\
\hline
$I_{15}$ $\sim$ $I_{24}$  &  0.952 & 0.000 \\
$I_{15}$ $\sim$ $I_{90}$  &  0.880 & 0.003 \\
$I_{24}$ $\sim$ $I_{90}$  &  0.905 & 0.002 \\
\hline
$I_{15}$ $\sim$ fuv1A1  &  0.514  &  0.191 \\
$I_{15}$ $\sim$ fuv1A2  &  0.071  &  0.866 \\
$I_{15}$ $\sim$ fuv1B1  &  0.523  &  0.182 \\
$I_{15}$ $\sim$ fuv1B2  &  0.452  &  0.260 \\
$I_{15}$ $\sim$ fuv2A1  &  0.523  &  0.182 \\
$I_{15}$ $\sim$ fuv2A2  &  0.523  &  0.182 \\
$I_{15}$ $\sim$ fuv2B1  &  0.452  &  0.260 \\
\hline
$I_{24}$ $\sim$ fuv1A1  &   0.431  &   0.286 \\
$I_{24}$ $\sim$ fuv1A2  &   0.000  &   1.000 \\
$I_{24}$ $\sim$ fuv1B1  &   0.452  &   0.260 \\
$I_{24}$ $\sim$ fuv1B2  &   0.428  &   0.289 \\
$I_{24}$ $\sim$ fuv2A1  &   0.500  &   0.207 \\
$I_{24}$ $\sim$ fuv2A2  &   0.452  &   0.260 \\
$I_{24}$ $\sim$ fuv2B1  &   0.428  &   0.289 \\
\hline
$I_{90}$ $\sim$ fuv1A1  &   0.419  &  0.301 \\
$I_{90}$ $\sim$ fuv1A2  &   0.047  &  0.910 \\
$I_{90}$ $\sim$ fuv1B1  &   0.452  &  0.260 \\
$I_{90}$ $\sim$ fuv1B2  &   0.404  &  0.319 \\
$I_{90}$ $\sim$ fuv2A1  &   0.500  &  0.207 \\
$I_{90}$ $\sim$ fuv2A2  &   0.452  &  0.260 \\
$I_{90}$ $\sim$ fuv2B1  &   0.404  &  0.319 \\
\hline
\end{tabular}
\end{table}
\end{center}

Stephens et al. (2014)[11] states that the PAH mass fraction increases significantly toward molecular clouds except when there is a very strong radiation field since PAHs are likely being destroyed in such a field. The PAH mass fraction increases as one leaves the central OB association. On the other hand, the VSG mass fraction increases at locations of an enhanced radiation field. Expanding bubbles may be launching dust at velocities that can cause big grains to shatter into VSGs causing 24 $\mu$m emission (Stephens et al. 2014)[11]. From Table \ref{Spitzer_all}, the \textit{Spitzer} correlations for all 15 locations taken together show that the emission at 8$\mu$m is better correlated to the FUV as compared to the emission at 24 $\mu$m contrary to our assumptions. This may be because these 15 locations contain a mixture of both hot and relatively cold regions of the LMC. Hot regions refer to the HII regions. The cold regions refer to the general diffuse ISM. This is supported by the fact that we get a different correlation trend once we separate the N11 HII region from the other locations (Table \ref{Spitzer_N11}). In this case we see that at each FUV wavelength, the 24 $\mu$m emission is marginally better correlated to the FUV as compared to the 8 $\mu$m emission. This supports the existing theory that the 24 $\mu$m VSG emission is seen to be associated with locations close to the hot UV emitting stars like HII regions, same as the dust scattered UV radiation. \\

The 24 $\mu$m emission is known to be dominated by warm dust emission from VSGs, which are mainly heated by young massive stars. Results by \textit{Spitzer} (Wu 2005[4]; Perez- Gonzalez et al. 2006[19]; Calzetti et al. 2007[20]) have shown that the 24 $\mu$m luminosity is one of the best Star Formation Rate (SFR) indicators. \textit{IRAS} and \textit{ISO} observations have proved that the FIR luminosity is also a good SFR tracer because of the emission peak around 60 $\mu$m of dust heated by star formation, thus the 70 $\mu$m emission must be closely related to star formation activities and should have tight and linear correlation with 24 $\mu$m warm dust emission (Zhu et al. 2008)[21]. As seen from Table \ref{Akari_all}, the \textit{AKARI} 15 $\mu$m, 24 $\mu$m and 90 $\mu$m show a good correlation with one another which shows they are associated with VSGs from similar hot environments. This becomes more prominent when we take into consideration the 30 Doradus HII region (Table \ref{Akari_30D}) which clearly gives us better correlation values among the 15 $\mu$m, 24 $\mu$m and 90 $\mu$m emissions. The correlations between 15 $\mu$m and FUV seem to be better in the HII 30 Doradus region as compared to when all hot and cold regions are considered together. Also, the 24 $\mu$m and 90 $\mu$m correlations with FUV are better when we consider all 28 \textit{AKARI} locations as compared to when we consider only 30 Doradus. The FUV/IR(90 $\mu$m) ratio is a measure of the optical depth of the medium. We see that the average FUV/IR(90 $\mu$m) value in 30 Doradus is $0.0475$ as compared to the average FUV/IR(90 $\mu$m) value in N11 which is lower at $0.01437$. Hence, the lower correlation values in the 30 Doradus region may be attributed to its higher FUV/IR(90 $\mu$m) ratio.

\section{Conclusions}

\begin{itemize} 

\item In this work, we compare the two regions N11 (Table \ref{Spitzer_N11}) and 30 Doradus (Table \ref{Akari_30D}) in the LMC and observe better FUV$\sim$IR correlations for N11 ($\sim$ 0.8) as compared to 30 Doradus ($\sim$ 0.5).

\item We observe higher FUV/IR(90 $\mu$m) ratio for 30 Doradus in comparison to N11, which may indicate low extinction and/or more number of stars being unaffected by interstellar dust. 

\item 30 Doradus is a very complex region with a high density of stars and therefore more starlight, as FUV can be contributed by unresolved stars.

\item There is also a possibility of destruction of VSGs (24 $\mu$m emission) in 30 Doradus. Thus lower emission at 24 $\mu$m band.

\end{itemize}

We will try to model both the regions theoretically by using suitable dust mixtures. For 30 Doradus one needs to use a special dust mixture and N11 can be modelled by using a Milky Way type of dust mixture. We plan to use the stellar/diffuse fraction which was presented by Pradhan et al. (2010)[17] and postulate if the stellar component of the FUV is higher in 30 Doradus or not.

\section*{Acknowledgements}
This work is based in part on observations made with the \textit{Spitzer Space Telescope}, which is operated by the Jet Propulsion Laboratory, California Institute of Technology under a contract with NASA. This work is based on observations with \textit{AKARI}, a JAXA project with the participation of ESA. This research has made use of the \textit{NED} and the \textit{SIMBAD} databases.\\

AP acknowledges financial support from SERB DST FAST TRACK grant (SERB/ F/5143/2013–2014) and support from the DST – JSPS grant (DST/INT/ JSPS/P-189/2014). RG and AP thank the Inter-University Centre for Astronomy and Astrophysics, Pune for associateship. PS would like to thank Tezpur University for their support and hospitality that allowed for completion of this work.
 
\section*{References}

\begin{enumerate}

\item R. J. Trumpler, Spectrophotometric Measures of Interstellar Light Absorption,
42 (1930) 267. doi:10.1086/124051

\item B. T. Draine, Interstellar Dust Grains, 41
(2003) 241–289. arXiv:astro-ph/0304489,
doi:10.1146/annurev.astro.41.011802.094840.

\item N. V. Sujatha, P. Shalima, J. Murthy, R. C. Henry, Dust Properties
in the Far-Ultraviolet in Ophiuchus, 633 (2005) 257–261.
arXiv:astro-ph/0507125, doi:10.1086/444532.

\item J. Wu, N. J. Evans, II, Y. Gao, P. M. Solomon, Y. L. Shirley, P. A. Vanden
Bout, Connecting Dense Gas Tracers of Star Formation in our Galaxy to
High-z Star Formation, 635 (2005) L173–L176. arXiv:astro-ph/0511424,
doi:10.1086/499623.

\item K. Murata, Y. Koyama, M. Tanaka, H. Matsuhara, T. Kodama, Environmental
dependence of polycyclic aromatic hydrocarbon emission at z ˜ 0.8.
Investigation by observing RX J0152.7-1357with AKARI, 581 (2015) A114.
arXiv:1507.00094, doi:10.1051/0004-6361/201526276.

\item M. W. Werner, K. I. Uchida, K. Sellgren, M. Marengo, K. D. Gordon,
P. W. Morris, J. R. Houck, J. A. Stansberry, New Infrared Emission
Features and Spectral Variations in NGC 7023, 154 (2004) 309–314.
arXiv:astro-ph/0407213, doi:10.1086/422413.

\item H. Murakami, H. Baba, P. Barthel, D. L. Clements, M. Cohen, Y. Doi,
K. Enya, E. Figueredo, N. Fujishiro, H. Fujiwara, M. Fujiwara, P. Garcia-
Lario, T. Goto, S. Hasegawa, Y. Hibi, T. Hirao, N. Hiromoto, S. S. Hong,
K. Imai, M. Ishigaki, M. Ishiguro, D. Ishihara, Y. Ita, W.-S. Jeong, K. S.
Jeong, H. Kaneda, H. Kataza, M. Kawada, T. Kawai, A. Kawamura,
M. F. Kessler, D. Kester, T. Kii, D. C. Kim, W. Kim, H. Kobayashi,
B. C. Koo, S. M. Kwon, H. M. Lee, R. Lorente, S. Makiuti, H. Matsuhara,
T. Matsumoto, H. Matsuo, S. Matsuura, T. G. M¨uller, N. Murakami,
H. Nagata, T. Nakagawa, T. Naoi, M. Narita, M. Noda, S. H.
Oh, A. Ohnishi, Y. Ohyama, Y. Okada, H. Okuda, S. Oliver, T. Onaka,
T. Ootsubo, S. Oyabu, S. Pak, Y.-S. Park, C. P. Pearson, M. Rowan-
Robinson, T. Saito, I. Sakon, A. Salama, S. Sato, R. S. Savage, S. Serjeant,
H. Shibai, M. Shirahata, J. Sohn, T. Suzuki, T. Takagi, H. Takahashi,
T. Tanab´e, T. T. Takeuchi, S. Takita, M. Thomson, K. Uemizu,
M. Ueno, F. Usui, E. Verdugo, T. Wada, L. Wang, T. Watabe,
H. Watarai, G. J. White, I. Yamamura, C. Yamauchi, A. Yasuda, The
Infrared Astronomical Mission AKARI, 59 (2007) 369. arXiv:0708.1796,
doi:10.1093/pasj/59.sp2.S369.

\item T. Onaka, H. Matsuhara, T. Wada, N. Fujishiro, H. Fujiwara, M. Ishigaki,
D. Ishihara, Y. Ita, H. Kataza, W. Kim, T. Matsumoto, H. Murakami,
Y. Ohyama, S. Oyabu, I. Sakon, T. Tanab´e, T. Takagi, K. Uemizu,
M. Ueno, F. Usui, H. Watarai, M. Cohen, K. Enya, T. Ootsubo,
C. P. Pearson, N. Takeyama, T. Yamamuro, Y. Ikeda, The Infrared Camera
(IRC) for AKARI – Design and Imaging Performance, 59 (2007) 401.
arXiv:0705.4144, doi:10.1093/pasj/59.sp2.S401.

\item B. E. J. Pagel, M. G. Edmunds, R. A. E. Fosbury, B. L. Webster, A survey
of chemical compositions of H II regions in the Magellanic Clouds, 184
(1978) 569–592.

\item M. Feast, The Distance to the Large Magellanic Cloud; A Critical Review,
in: Y.-H. Chu, N. Suntzeff, J. Hesser, D. Bohlender (Eds.), New Views of
the Magellanic Clouds, Vol. 190 of IAU Symposium, 1999, p. 542.

\item I. W. Stephens, J. M. Evans, R. Xue, Y.-H. Chu, R. A. Gruendl,
D. M. Segura-Cox, Spitzer Observations of Dust Emission from H II Regions
in the Large Magellanic Cloud, 784 (2014) 147. arXiv:1402.2631,
doi:10.1088/0004-637X/784/2/147.

\item M. E. Putman, B. K. Gibson, L. Staveley-Smith, G. Banks, D. G. Barnes,
R. Bhatal, M. J. Disney, R. D. Ekers, K. C. Freeman, R. F. Haynes, P. Henning,
H. Jerjen, V. Kilborn, B. Koribalski, P. Knezek, D. F. Malin, J. R.
Mould, T. Oosterloo, R. M. Price, S. D. Ryder, E. M. Sadler, I. Stewart,
F. Stootman, R. A. Vaile, R. L. Webster, A. E. Wright, Tidal disruption
of the Magellanic Clouds by the Milky Way, 394 (1998) 752–754.
arXiv:astro-ph/9808023, doi:10.1038/29466.

\item V. Gorjian, M. W. Werner, J. R. Mould, K. D. Gordon, J. Muzzerole,
J. Morrison, J. M. Surace, L. M. Rebull, R. L. Hurt, R. C.
Smith, S. D. Points, C. Aguilera, J. M. D. Buizer, C. Packham,
Infrared imaging of the large magellanic cloud star-forming region henize 206,
The Astrophysical Journal Supplement Series 154 (1) (2004) 275.
URL http://stacks.iop.org/0067-0049/154/i=1/a=275

\item K. G. Henize, Catalogues of H-EMISSION Stars and Nebulae in the Magellanic
Clouds., 2 (1956) 315. doi:10.1086/190025.

\item M. Meixner, K. D. Gordon, R. Indebetouw, J. L. Hora, B. Whitney,
R. Blum, W. Reach, J.-P. Bernard, M. Meade, B. Babler, C. W. Engelbracht,
B.-Q. For, K. Misselt, U. Vijh, C. Leitherer, M. Cohen, E. B.
Churchwell, F. Boulanger, J. A. Frogel, Y. Fukui, J. Gallagher, V. Gorjian,
J. Harris, D. Kelly, A. Kawamura, S. Kim, W. B. Latter, S. Madden,
C. Markwick-Kemper, A. Mizuno, N. Mizuno, J. Mould, A. Nota, M. S.
Oey, K. Olsen, T. Onishi, R. Paladini, N. Panagia, P. Perez-Gonzalez,
H. Shibai, S. Sato, L. Smith, L. Staveley-Smith, A. G. G. M. Tielens,
T. Ueta, S. van Dyk, K. Volk, M. Werner, D. Zaritsky, Spitzer Survey of
the Large Magellanic Cloud: Surveying the Agents of a Galaxy’s Evolution
(SAGE). I. Overview and Initial Results, 132 (2006) 2268–2288.
arXiv:astro-ph/0606356, doi:10.1086/508185.

\item M. O. Oestreicher, T. Schmidt-Kaler, The dust distribution inside the Large
Magellanic Cloud., 117 (1996) 303–312.

\item A. C. Pradhan, A. Pathak, J. Murthy,
Far-ultraviolet diffuse emission from the large magellanic cloud, The
Astrophysical Journal Letters 718 (2) (2010) L141.
URL http://stacks.iop.org/2041-8205/718/i=2/a=L141

\item J. Murthy, D. J. Sahnow, Observations of the Diffuse Far-Ultraviolet Background
with the Far Ultraviolet Spectroscopic Explorer, 615 (2004) 315–
322. arXiv:astro-ph/0407519, doi:10.1086/424441

\item P. G. P´erez-Gonz´alez, R. C. Kennicutt, Jr., K. D. Gordon, K. A. Misselt,
A. Gil de Paz, C. W. Engelbracht, G. H. Rieke, G. J. Bendo, L. Bianchi,
S. Boissier, D. Calzetti, D. A. Dale, B. T. Draine, T. H. Jarrett, D. Hollenbach,
M. K. M. Prescott, Ultraviolet through Far-Infrared Spatially Resolved
Analysis of the Recent Star Formation in M81 (NGC 3031), 648
(2006) 987–1006. arXiv:astro-ph/0605605, doi:10.1086/506196.

\item D. Calzetti, R. C. Kennicutt, Jr., L. Bianchi, D. A. Thilker, D. A. Dale,
C. W. Engelbracht, C. Leitherer, M. J. Meyer, M. L. Sosey, M. Mutchler, M. W. Regan, M. D. Thornley, L. Armus, G. J. Bendo, S. Boissier,
A. Boselli, B. T. Draine, K. D. Gordon, G. Helou, D. J. Hollenbach, L. Kewley,
B. F. Madore, D. C. Martin, E. J. Murphy, G. H. Rieke, M. J. Rieke,
H. Roussel, K. Sheth, J. D. Smith, F. Walter, B. A. White, S. Yi, N. Z.
Scoville, M. Polletta, D. Lindler, Star Formation in NGC 5194 (M51a):
The Panchromatic View from GALEX to Spitzer, 633 (2005) 871–893.
arXiv:astro-ph/0507427, doi:10.1086/466518.

\item Y.-N. Zhu, H. Wu, C. Cao, H.-N. Li, Correlations between Mid-Infrared, Far-Infrared, H$\alpha$, and FUV Luminosities for Spitzer SWIRE Field Galaxies,
686 (2008) 155–171. doi:10.1086/591121.

\end{enumerate}

\end{document}